\documentstyle[11pt,newpasp,twoside,epsf]{article}

\begin{document}

\title{The Munich Near-IR Cluster Survey (MUNICS)}
\author{N. Drory$^1$, U. Hopp$^1$, R. Bender$^1$, G. Feulner$^1$, 
J. Snigula$^1$,
C. Mendes de Oliveira$^2$, G. Hill$^3$}
\affil{$^1$Universit\"ats-Sternwarte, Scheinerstr. 1, 81679 M\"unchen, Germany}
\affil{$^3$Instituto Astron\^{o}mico e Geof\'{\i}sico, S\~{a}o Paulo, Brazil}
\affil{$^2$University of Texas at Austin, Austin, Texas 78712}

\begin{abstract}
The Munich Near-IR Cluster Survey (MUNICS) is a $K'$ selected survey
covering 1 square degree in the $K'$ and $J$ NIR bands with
complementary optical photometry in the $V$, $R$, and $I$ bands
covering a subarea of 0.35 square degrees. The $3\sigma$ limiting
magnitude is 19.5 in $K'$. The main goals of the project are the
identification of clusters of galaxies at redshifts $0.6<z<1.5$ and
the study of the evolution of the early-type field population at
similar redshifts. Here we present first results regarding color
distributions and the surface densities of EROs as well as photometric
redshifts and a first clustering analysis of the sample.
\end{abstract}

\section{Introduction}
The MUNICS project is a wide area $K'$-band selected photometric
survey in the $VRIJK'$ passbands aiming at two main scientific goals,
namely
\begin{itemize}
\item the identification of galaxy clusters at redshifts around unity, and
\item the selection of a fair sample of field early-type galaxies at
similar redshifts for evolutionary studies.
\end{itemize}
Near-IR selection is an efficient tool for tracing the massive galaxy
population at redshifts around unity because of its high sensitivity
for evolved stellar populations even in the presence of moderate star
formation activity.  Thus a $K$-band selected survey can provide a
very useful database for the investigation of the formation and
evolution of the cluster as well as the field population of massive
galaxies.
\begin{figure}[ht!]
\plottwo{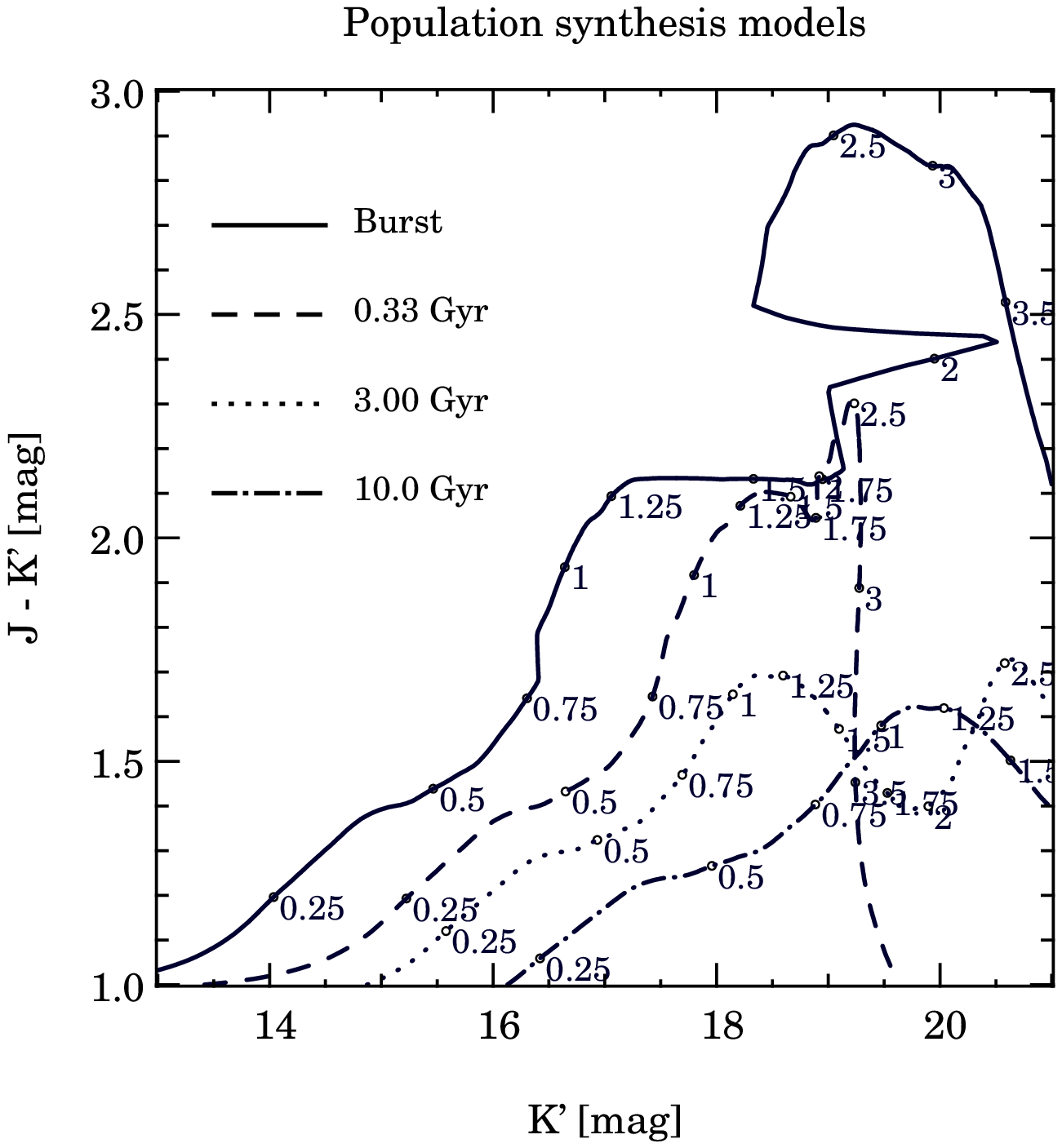}{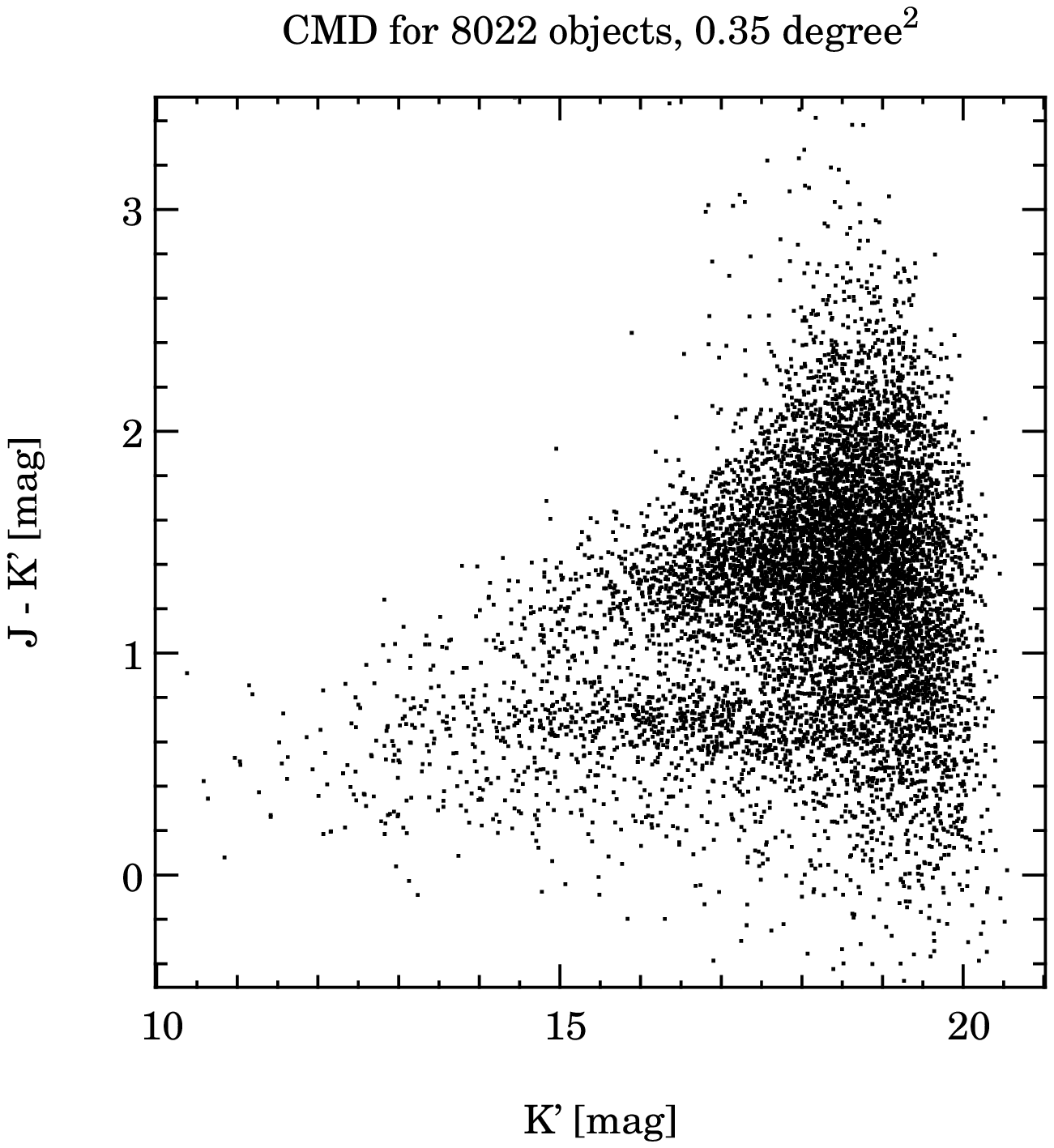}
\caption{{\em Left panel:} $K'$ vs. $J\!-\!K'$ color-magnitude diagram
for population synthesis models for a simple stellar population and 3
exponential star formation histories with timescales given in the
legend. The galaxies are assumed to form at $z_f = 4$ and to have a
luminosity of $L_*$ today. {\em Right panel:} the same diagram for
0.35 square degrees of the MUNICS data. The sequence at a color $\leq 0.8$
are late-type stars.}
\end{figure}

Clusters of galaxies are of prime interest in extragalactic astronomy
and cosmology.  The $z$ evolution of their number density and
correlation function are sensitive tests for structure formation
theories and especially the density parameter $\Omega$. Models of
structure formation predict that if $\Omega_0 = 1$, the number density
of clusters of richness class 1 declines by a factor of $10^3$ between
$z=0$ and $z=1$. On the other hand, if $\Omega_0 = 0.3$, the number
density declines only by a factor of $\leq 10$ in the same redshift
range (Eke, Cole, \& Frenk 1996; Bahcall, Fan, \& Cen 1997; Fan,
Bahcall, \& Cen 1997).  Furthermore, clusters of galaxies allow to
find large numbers of massive galaxies at higher redshift and thus represent
unique laboratories to study the evolution of galaxies in high density
regions as a function of redshift. While the number of clusters known
at redshifts $z>0.5$ is steadily increasing, {\em uniformly selected}
samples of clusters at high redshift are still deficient in the
optical and near-IR wavelength ranges.

The formation and evolution of the population of massive galaxies is
still a matter of lively and controversial debate. No general
agreement has been reached yet regarding the formation era of
spheroidals. While models of hierarchical galaxy formation (Cole et
al. 1993; Kauffman \& Charlot 1998) consistently predict a steep
decline in the number density of massive spheroidals, they have a
rather large number of free parameters, some of which involve
ill-understood processes. Observation has not yet been successful in
constraining the ranges of the involved model parameters tightly
enough, so that comparisons between theory and experiment are difficult
to interpret. Moreover, measuring the evolution of the number density
of early-type galaxies $\partial N(z)/\partial z$ to redshifts of
unity is by itself a difficult undertaking, suffering from too small
samples and strong selection effects, therefore yielding contradictory
results (e.g. Totani \& Yoshii 1998; Benitez et al. 1999; Shade et
al. 1999; Broadhurst \& Bouwens 1999; Barger et al. 1999).

\section{The MUNICS Project}
\begin{figure}[ht!]
\begin{center}
\plottwo{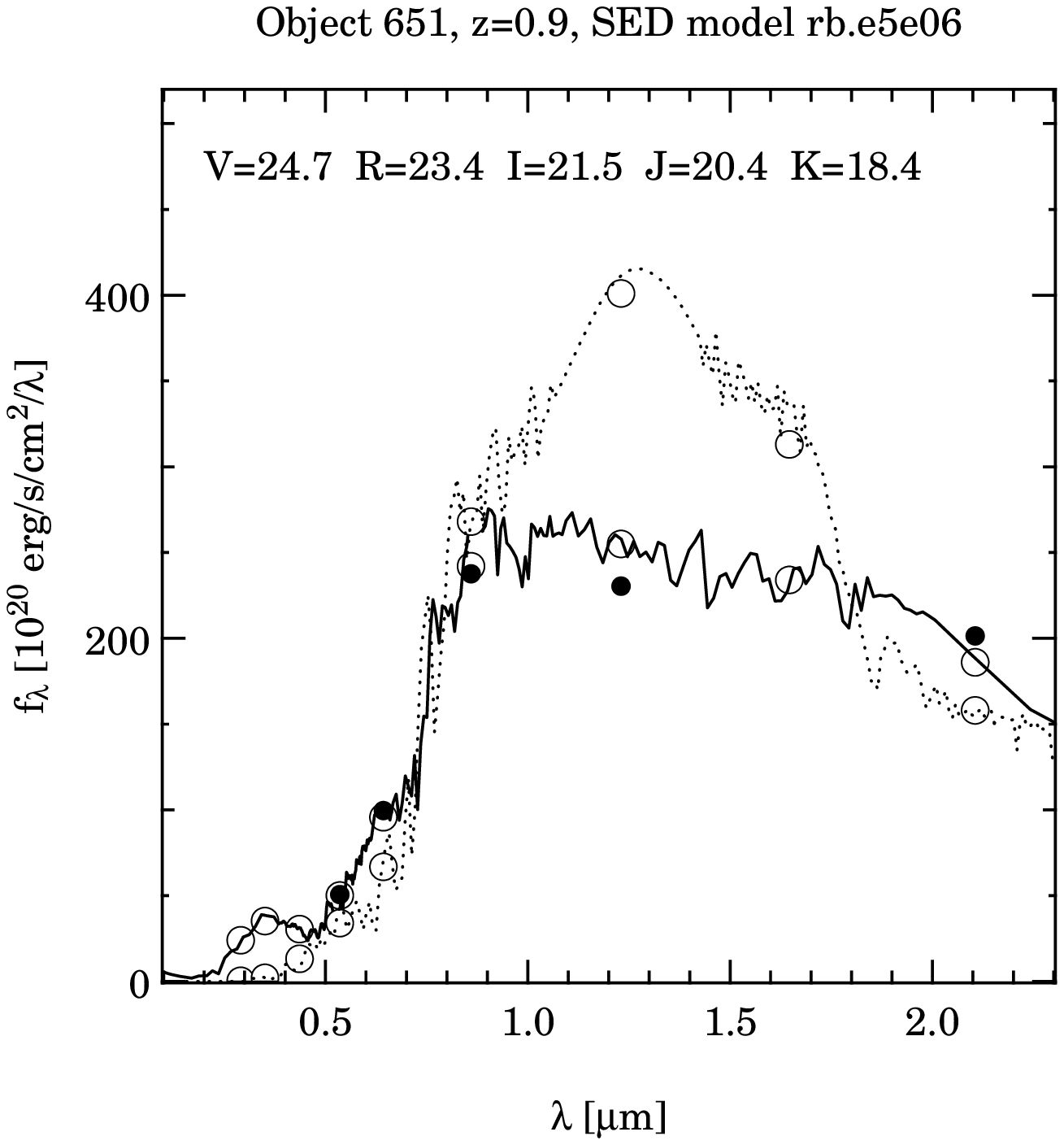}{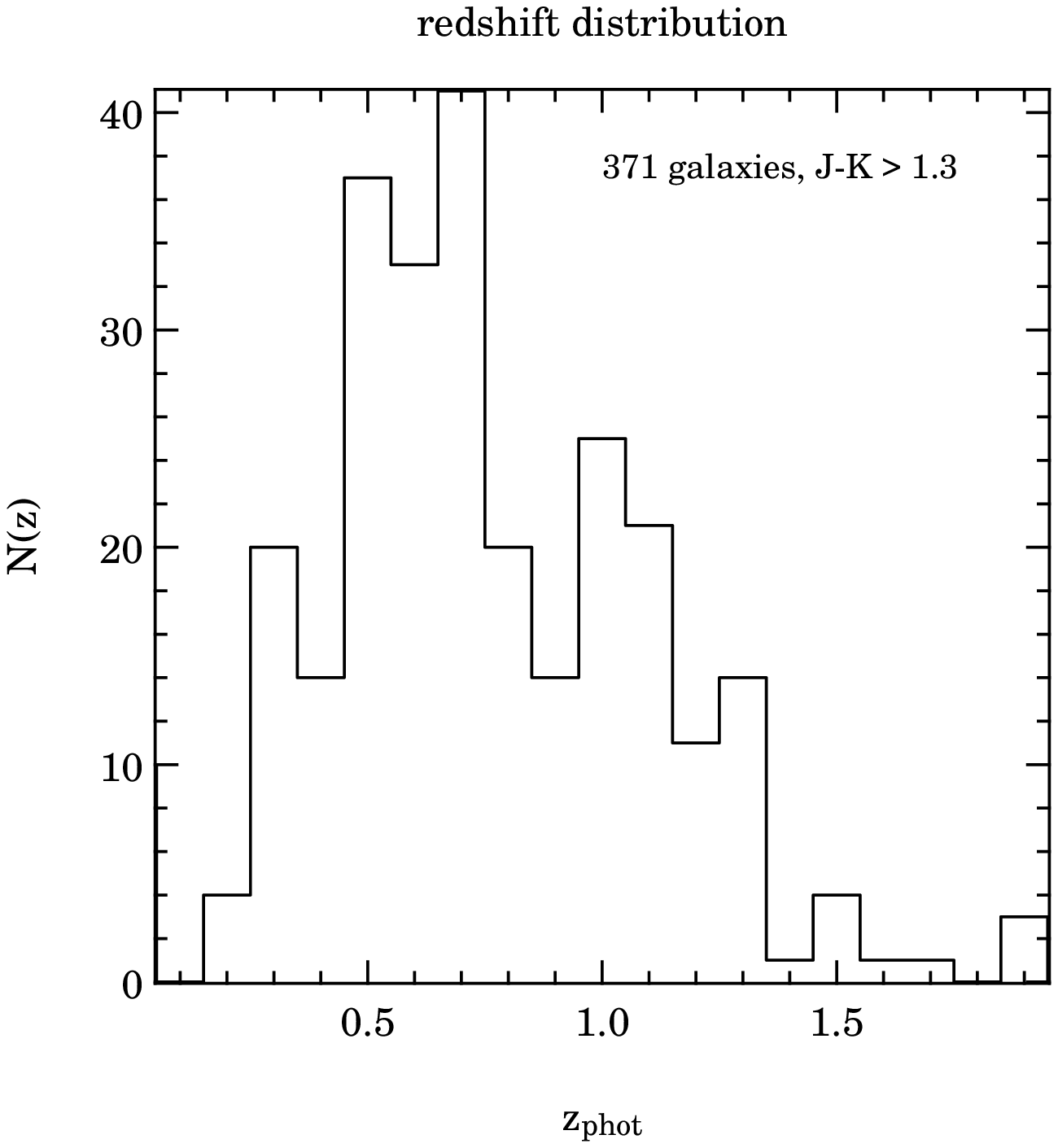}
\end{center}
\caption{{\em Left panel:} Magnitudes and best fit SED for an object
at $z=0.9$. Black dots denote the measured fluxes in $VRIJK'$, the
solid line is the best fit redshifted galaxy SED used to derive the
photometric redshift, a 5 Gyr old `elliptical' in this case. The dashed
line is the best fit local stellar SED, a M5 type star. Note that the
$J$-band flux is the only distinguishing feature in such a
case. {\em Right panel:} photometric redshift distributon for 371
galaxies selected out of a $13.2\arcmin\times 13.2\arcmin$ field.}
\end{figure}
The MUNICS project uniformly covers 1 square degree in the $J$ and
$K'$ near-IR bands. The survey area consists of 8 $13.2\arcmin \times
26.2\arcmin$ randomly selected fields at high galactic latitude, as
well as 13 $7\arcmin \times 7\arcmin$ fields targeted towards
$0.6<z<1.5$ QSOs. The $3\sigma$ detection limits for a point source
are $19.5$ in the $K'$-band and $21.5$ in the $J$-band. The data have
been acquired at the 3.5m telescope at Calar Alto Observatory using
the $\Omega\!-\!{\rm Prime}$ camera.

Optical photometry in the $V$, $R$, and $I$ bands was obtained
for a subsample of the survey fields covering 0.35 square degrees in
total. These data have been obtained at the 2.2m telescope at Calar
Alto Observatory and the 2.7m telescope at McDonald Observatory. These
data will enable us to determine photometric redshifts for the
galaxies and thus are of great importance in selecting and confirming
cluster candidates as well as individual galaxies for follow-up
spectroscopy.

Figure 1 shows $K'$ vs. $J\!-\!K'$ color magnitude diagrams for
population synthesis models based on the Bruzual \& Charlot 1995 code
(Bruzual \& Charlot 2000), together with a
subsample of the MUNICS data. Note that the $J\!-\!K'$ color of a late
M-type star is $\leq 0.8$, and therefore any object having redder
color must be redshifted, with the exception of extreme stellar
objects like brown dwarfs. Those are not expected to be very numerous
down to our sensitivity limits. Therefore the $J\!-\!K'$ color is a
very powerful selection tool for picking out the evolved populations
of massive galaxies at redshifts around unity and larger.

Figure 2 (right panel) shows the photometric redshift distribution of
371 galaxies which were detected in {\em all} 5 passbands in a small
subarea of $13.2\arcmin \times 13.2\arcmin$ and have a $J\!-\!K'$
color $\geq 1.3$, confirming the usefulness of the $J\!-\!K'$
selection. The left panel demonstrates that we can reliably
reconstruct the SED of early-type galaxies at high redshift from our
$VRIJK$ data.

Data reduction and calibration of the near-IR data is completed and
reduction of the optical data is almost completed. Here we present
first results regarding detection of clusters and the surface density
of EROs.

\section{The Surface Density of EROs}
The surface density of Extremely Red Objetcs (EROs) defined in terms
of $J\!-\!K'$ as determined from 0.35 square degrees of data is given
in Table 1. As soon as the optical data become fully available further
investigation of the nature of such objects will be possible, as well
as comparisons to previous studies which mostly use $R\!-\!K$ or
$I\!-\!K$ for defining EROs, and have surveyed much smaller areas at
comparable depth.

\begin{table}
\begin{center}
\begin{tabular}[ht!]{c|r|l}
\hline
	$J\!-\!K'$ & $N_{total}$ & $\frac{N}{arcmin^2}$ \\
\hline
\hline
	$> 1.50$   &  3030  & 2.41   $\pm$ 0.04   \\
	$> 1.75$   &  1521  & 1.21   $\pm$ 0.03   \\
	$> 2.00$   &  706   & 0.56   $\pm$ 0.021  \\
	$> 2.25$   &  301   & 0.24   $\pm$ 0.014  \\
	$> 2.50$   &  124   & 0.098  $\pm$ 0.009  \\
	$> 2.75$   &  51    & 0.040  $\pm$ 0.006  \\
	$> 3.00$   &  27    & 0.021  $\pm$ 0.004  \\
	$> 3.25$   &  11    & 0.009  $\pm$ 0.003  \\
	$> 3.50$   &  5     & 0.004  $\pm$ 0.001  \\
\hline
\end{tabular}
\caption{Surface densities as a function of $J\!-\!K'$ color as derived
from a total of 0.35 square degrees.}
\end{center}
\end{table}
Assuming that the $J\!-\!K'$ and $R\!-\!K'$ colors for LBDS 53W091
(Spinrad et al. 1997) are typical for EROs, all objects with $J\!-\!K'
> 1.75$ have to be considered as ERO candidates. The values of Table 1
then point to higher surface densities than the values obtained by
Thompson et al. (1999), which were based on an $R\!-\!K' > 6$ color
and a survey area of 0.04 square degrees.

\section{Detection of Galaxy Clusters}
In recent years it became clear that an early type population in
clusters was well in place at redshifts of at least 0.8
(e.g. Stanford, Eisenhardt, \& Dickinson 1998). Thus we may hope to
detect these clusters by looking for overdensities of red objects with
colors resembling the color sequence of cluster early-type
galaxies. The surface density field is divided into (overlapping)
slices in $J\!-\!K'$ color. These slices are then smoothed by a kernel
of the angular size of a cluster core at the apropriate redshift for
that particular color range. Cluster candidates are identified as
overdensities in this data cube.

Figure 3 shows the redshift distribution for objects in the vicinity
of a high redshift cluster candidate detected in this fashion as a
demonstration of the efficiency of the technique. The cluster
candidate is first detected as an overdensity of red objects in
$J\!-\!K'$. Then this detection is verified by looking at the
histogram of photometric redshifts in the vicinity of the cluster
candidate.
\begin{figure}[ht!]
\begin{center}
\begin{minipage}{0.4\textwidth}
\plotone{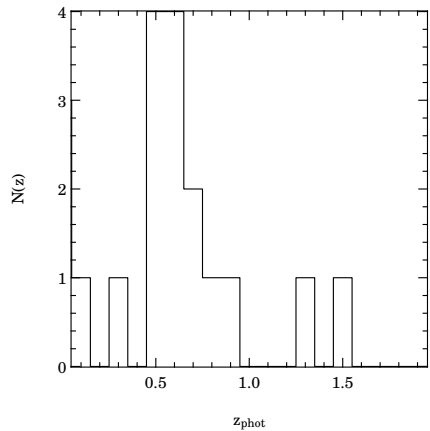}
\end{minipage}
\end{center}
\caption{Photometric redshift distribution of objects in the vicinity of a cluster candidate detected as an overdensity of red objects as described in the text. The cluster has an estimated redshift of 0.6.}
\end{figure}

\end{document}